\RequirePackage{lineno} 
\documentclass[aps,prl,twocolumn,showpacs,superscriptaddress]{revtex4}
\usepackage{graphicx,epsf,epsfig,color}
\usepackage{amssymb, bm}
\usepackage{lineno}

\newcommand{\dzero}{\mbox{$D^0$}}

\newcommand{\sqrtsNN}{\mbox{$\sqrt{s_{_{\mathrm{NN}}}}$}}
\newcommand{\AuAu}{Au+Au}
\newcommand{\pp}{\mbox{\textit{p}+\textit{p}}}
\newcommand{\pt}{\mbox{$p_{\rm T}$}}
\newcommand{\gevc}{\mbox{${\mathrm{GeV/}}c$}}
\newcommand{\RAA}{\mbox{$R_{\rm AA}$}}
\newcommand{\nbin}{\mbox{$N_{\mathrm{bin}}$}}
\newcommand{\npart}{\mbox{$N_{\mathrm{part}}$}}

\def\Journal#1#2#3#4{{#1} {\bf #2}, #3 (#4)}

\def\PRL{{\em Phys. Rev. Lett.}}
\def\PRD{{\em Phys. Rev.} D}
\def\PRC{{\em Phys. Rev.} C}
\def\EPJC{{\em Eur. Phys. J.} C}

\def\JSP{{\em J. Stat. Phys.}}

\begin{document}


\title{Erratum: Observation of $D^0$ meson nuclear modifications in Au+Au collisions at \sqrtsNN = 200\,GeV [Phys.~Rev.~Lett.~113,142301 (2014)]}
    
\date{DRAFT \today}

\affiliation{Abilene Christian University, Abilene, Texas   79699}
\affiliation{AGH University of Science and Technology, FPACS, Cracow 30-059, Poland}
\affiliation{Alikhanov Institute for Theoretical and Experimental Physics, Moscow 117218, Russia}
\affiliation{Argonne National Laboratory, Argonne, Illinois 60439}
\affiliation{Brookhaven National Laboratory, Upton, New York 11973}
\affiliation{University of California, Berkeley, California 94720}
\affiliation{University of California, Davis, California 95616}
\affiliation{University of California, Los Angeles, California 90095}
\affiliation{University of California, Riverside, California 92521}
\affiliation{Central China Normal University, Wuhan, Hubei 430079 }
\affiliation{University of Illinois at Chicago, Chicago, Illinois 60607}
\affiliation{Creighton University, Omaha, Nebraska 68178}
\affiliation{Czech Technical University in Prague, FNSPE, Prague 115 19, Czech Republic}
\affiliation{Technische Universit\"at Darmstadt, Darmstadt 64289, Germany}
\affiliation{E\"otv\"os Lor\'and University, Budapest, Hungary H-1117}
\affiliation{Frankfurt Institute for Advanced Studies FIAS, Frankfurt 60438, Germany}
\affiliation{Fudan University, Shanghai, 200433 }
\affiliation{University of Heidelberg, Heidelberg 69120, Germany }
\affiliation{University of Houston, Houston, Texas 77204}
\affiliation{Indiana University, Bloomington, Indiana 47408}
\affiliation{Institute of Modern Physics, Chinese Academy of Sciences, Lanzhou, Gansu 730000 }
\affiliation{Institute of Physics, Bhubaneswar 751005, India}
\affiliation{University of Jammu, Jammu 180001, India}
\affiliation{Joint Institute for Nuclear Research, Dubna 141 980, Russia}
\affiliation{Kent State University, Kent, Ohio 44242}
\affiliation{University of Kentucky, Lexington, Kentucky 40506-0055}
\affiliation{Lawrence Berkeley National Laboratory, Berkeley, California 94720}
\affiliation{Lehigh University, Bethlehem, Pennsylvania 18015}
\affiliation{Max-Planck-Institut f\"ur Physik, Munich 80805, Germany}
\affiliation{Michigan State University, East Lansing, Michigan 48824}
\affiliation{National Research Nuclear University MEPhI, Moscow 115409, Russia}
\affiliation{National Institute of Science Education and Research, HBNI, Jatni 752050, India}
\affiliation{National Cheng Kung University, Tainan 70101 }
\affiliation{Nuclear Physics Institute of the CAS, Rez 250 68, Czech Republic}
\affiliation{Ohio State University, Columbus, Ohio 43210}
\affiliation{Institute of Nuclear Physics PAN, Cracow 31-342, Poland}
\affiliation{Panjab University, Chandigarh 160014, India}
\affiliation{Pennsylvania State University, University Park, Pennsylvania 16802}
\affiliation{Institute of High Energy Physics, Protvino 142281, Russia}
\affiliation{Purdue University, West Lafayette, Indiana 47907}
\affiliation{Pusan National University, Pusan 46241, Korea}
\affiliation{Rice University, Houston, Texas 77251}
\affiliation{Rutgers University, Piscataway, New Jersey 08854}
\affiliation{Universidade de S\~ao Paulo, S\~ao Paulo, Brazil 05314-970}
\affiliation{University of Science and Technology of China, Hefei, Anhui 230026}
\affiliation{Shandong University, Qingdao, Shandong 266237}
\affiliation{Shanghai Institute of Applied Physics, Chinese Academy of Sciences, Shanghai 201800}
\affiliation{Southern Connecticut State University, New Haven, Connecticut 06515}
\affiliation{State University of New York, Stony Brook, New York 11794}
\affiliation{Temple University, Philadelphia, Pennsylvania 19122}
\affiliation{Texas A\&M University, College Station, Texas 77843}
\affiliation{University of Texas, Austin, Texas 78712}
\affiliation{Tsinghua University, Beijing 100084}
\affiliation{University of Tsukuba, Tsukuba, Ibaraki 305-8571, Japan}
\affiliation{United States Naval Academy, Annapolis, Maryland 21402}
\affiliation{Valparaiso University, Valparaiso, Indiana 46383}
\affiliation{Variable Energy Cyclotron Centre, Kolkata 700064, India}
\affiliation{Warsaw University of Technology, Warsaw 00-661, Poland}
\affiliation{Wayne State University, Detroit, Michigan 48201}
\affiliation{Yale University, New Haven, Connecticut 06520}

\author{J.~Adam}\affiliation{Creighton University, Omaha, Nebraska 68178}
\author{L.~Adamczyk}\affiliation{AGH University of Science and Technology, FPACS, Cracow 30-059, Poland}
\author{J.~R.~Adams}\affiliation{Ohio State University, Columbus, Ohio 43210}
\author{J.~K.~Adkins}\affiliation{University of Kentucky, Lexington, Kentucky 40506-0055}
\author{G.~Agakishiev}\affiliation{Joint Institute for Nuclear Research, Dubna 141 980, Russia}
\author{M.~M.~Aggarwal}\affiliation{Panjab University, Chandigarh 160014, India}
\author{Z.~Ahammed}\affiliation{Variable Energy Cyclotron Centre, Kolkata 700064, India}
\author{I.~Alekseev}\affiliation{Alikhanov Institute for Theoretical and Experimental Physics, Moscow 117218, Russia}\affiliation{National Research Nuclear University MEPhI, Moscow 115409, Russia}
\author{D.~M.~Anderson}\affiliation{Texas A\&M University, College Station, Texas 77843}
\author{R.~Aoyama}\affiliation{University of Tsukuba, Tsukuba, Ibaraki 305-8571, Japan}
% ...

\collaboration{STAR Collaboration}\noaffiliation

\pacs{25.75.-q}

\maketitle

~\\

In this erratum we report changes on the \dzero\ \pt\ spectra and nuclear modification factor (\RAA) in Au+Au collisions at \sqrtsNN\ = 200 GeV by fixing the errors in the efficiency and selection criteria that affected the Au+Au results. The p+p reference spectrum has changed as well and is updated with new fragmentation parameters.

\begin{figure}[tb]
\newcounter{TempEqCnt}
\setcounter{TempEqCnt}{\value{figure}}
\setcounter{figure}{1} 
\includegraphics[width=0.45\textwidth]{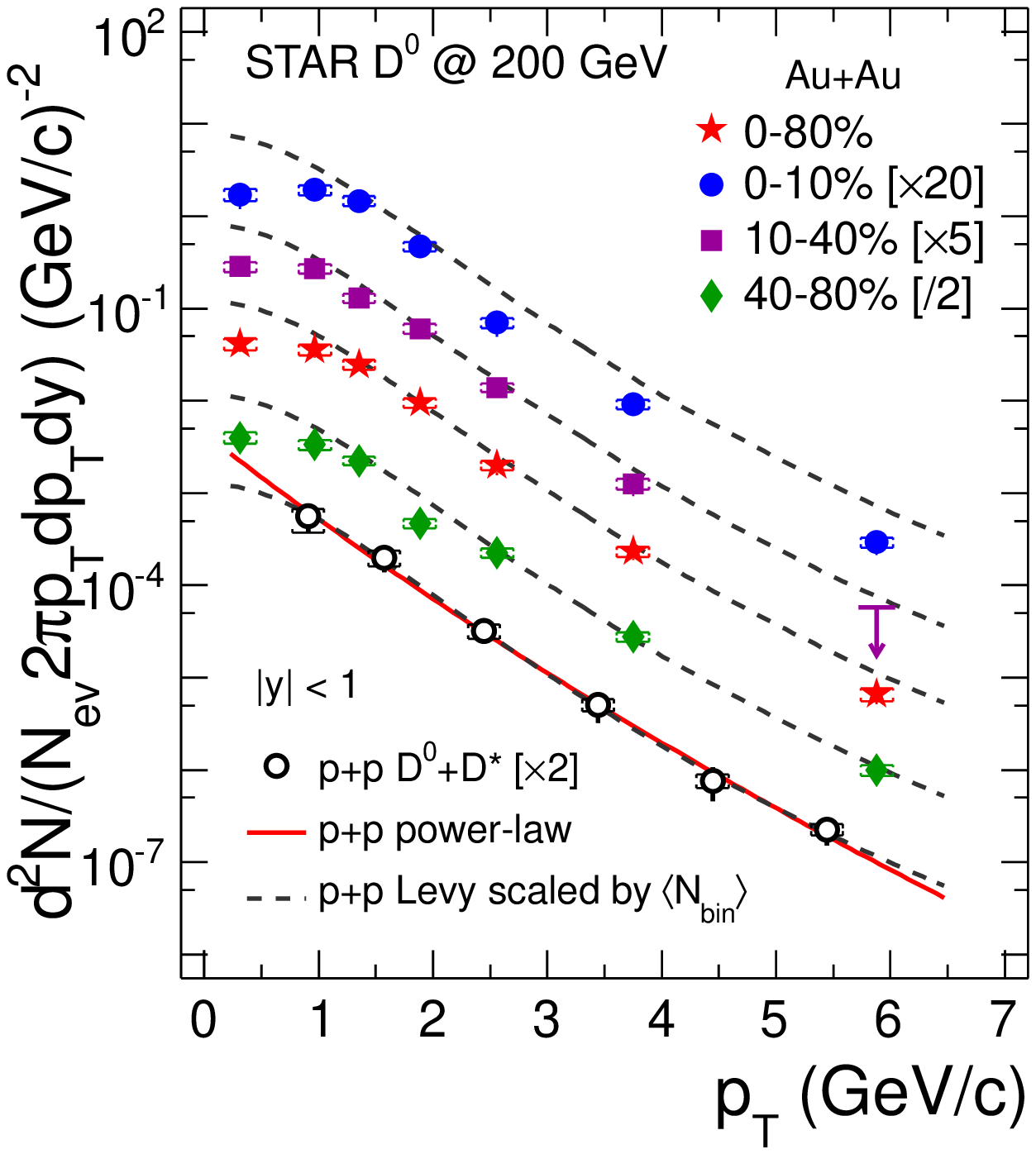}
\caption{(Color online) Centrality dependence of the \dzero\ \pt\ differential invariant yield in \AuAu\ collisions (solid symbols). The curves are number-of-binary-collision-scaled Levy functions from fitting to the \pp\ result (open circles), which has been updated from~\cite{ppCharm} with the latest global analysis of charm fragmentation ratios from Ref.~\cite{Charmfrag} and also taking into account the \pt\ dependence of the fragmentation ratio between \dzero\ and $D^{*{\pm}}$ from PYTHIA 6.4~\cite{PY6}. The arrow denotes the upper limit with 90\% confidence level of the last data point for 10$-$40\% collisions. The systematic uncertainties are shown as square brackets.} 
\label{fig:figure1}
\end{figure}

In the original Letter~\cite{D0PRL}, we reported on measurements of the nuclear modification factor of $D^0$ mesons in \AuAu\ collisions at \sqrtsNN\ = 200 GeV. We have identified two errors in the estimation of the particle identification efficiency. Considering the high combinatorial background in $D^0$ meson reconstruction in Au+Au collisions, a hybrid particle identification method was used in this analysis to improve significance of the signal.~
For $p < 1.6$ GeV/$c$, pion and kaon candidates were selected by requiring a selection on the ionization energy loss of the particle passing through the Time Projection Chamber (TPC), good matching to a hit on the Time of Flight detector (TOF), and a TOF $1/\beta$ (the reciprocal of particle velocity) selection. 
For $p > 1.6$ GeV/$c$, candidates were required to pass the TPC ionization energy loss ($dE/dx$) selection and the $1/\beta$ selection was required for those tracks with good TOF matching. This helps to enhance pion and kaon identification purity.
The first error was that in the analysis of the reconstructed data, we did not correctly reject tracks with TOF matching, but with no valid $\beta$ information due to unavailable calibration parameters. The second error was that we accounted for the efficiency of a distance of closest approach to primary vertex (DCA) selection twice. The DCA-in-the-transverse-plane selection was applied to tracks to insure a good TOF path length calculation by esuring the tracks are primary. The difference in efficiency from the two errors combined is 30\% for single tracks at low $p_T$ and it decreases with increasing $p_T$ compared with the previous result in Au+Au collisions. This results in a factor of two higher $D^0$ yields estimation at $p_T~<$ 2 GeV/$c$ compared to the case when the correct algorithm is used, affecting results in all centralities in~\cite{D0PRL}. 


\begin{figure}[tb]
\includegraphics[width=0.45\textwidth]{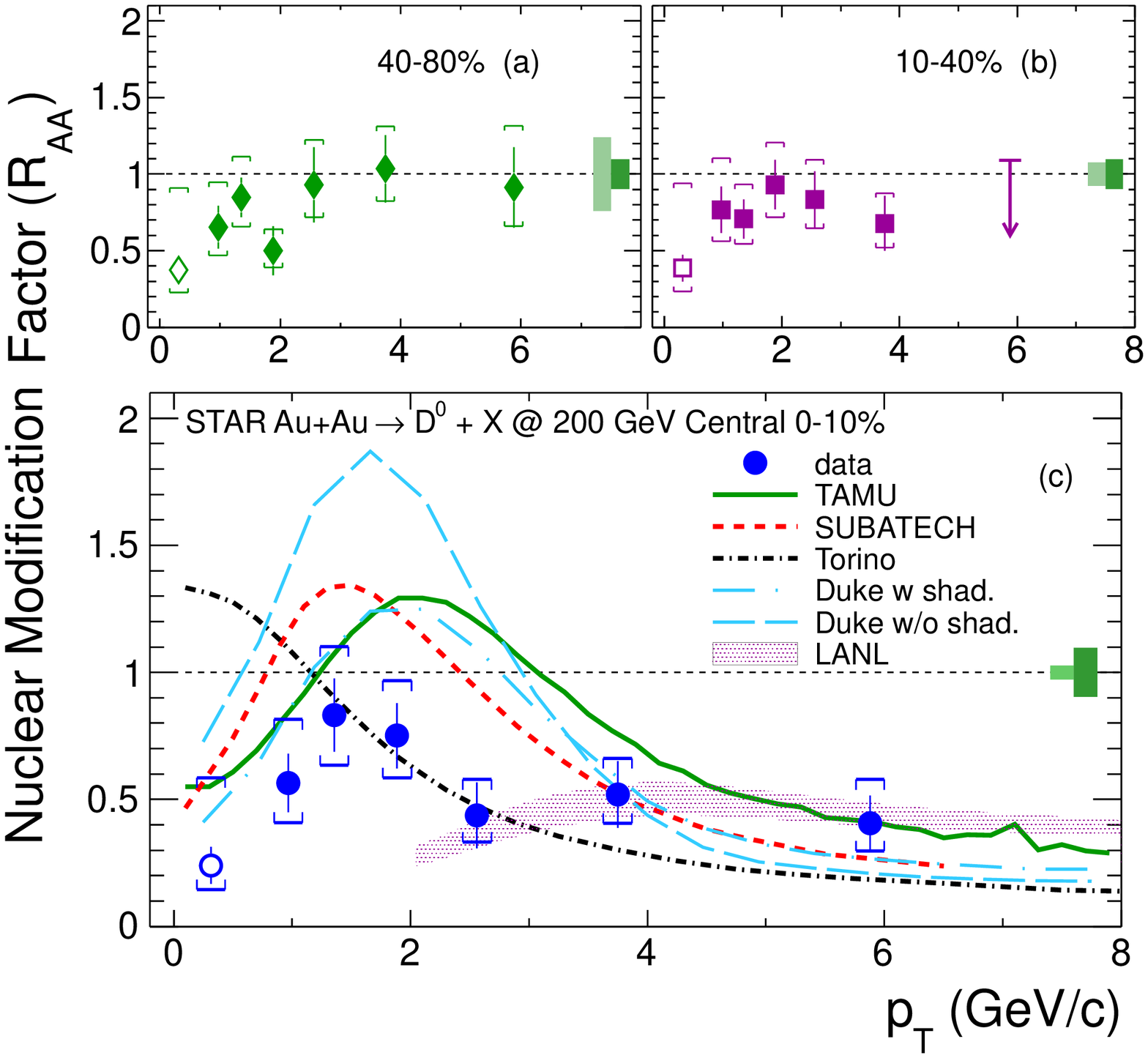}
\caption{(Color online) Panels (a)(b): \dzero\ \RAA\ for peripheral 40$-$80\% and semi-central 10$-$40\% collisions; Panel (c): \dzero\ \RAA\ for 0$-$10\% most central events (blue circles) compared with model calculations from the TAMU (solid curve), SUBATECH (dashed curve), Torino (dot-dashed curve), Duke (long-dashed and long-dot-dashed curves), and LANL groups (filled band). The open symbol indicates the result with the extrapolated \pp\ reference. The vertical lines and brackets around the data points denote the statistical and systematic uncertainties respectively. The vertical bars around unity denote the overall normalization uncertainties in the \AuAu\ and \pp\ data, respectively. The \RAA\ probability distribution for the 0$-$0.7 GeV/$c$ data point is largely skewed. The uncertainty we report is the 68.3\% probability range with respect to the measured central value assuming Gaussian distribution. }
\label{fig:figure2}
\end{figure}

\begin{figure}[tb]
\includegraphics[width=0.4\textwidth]{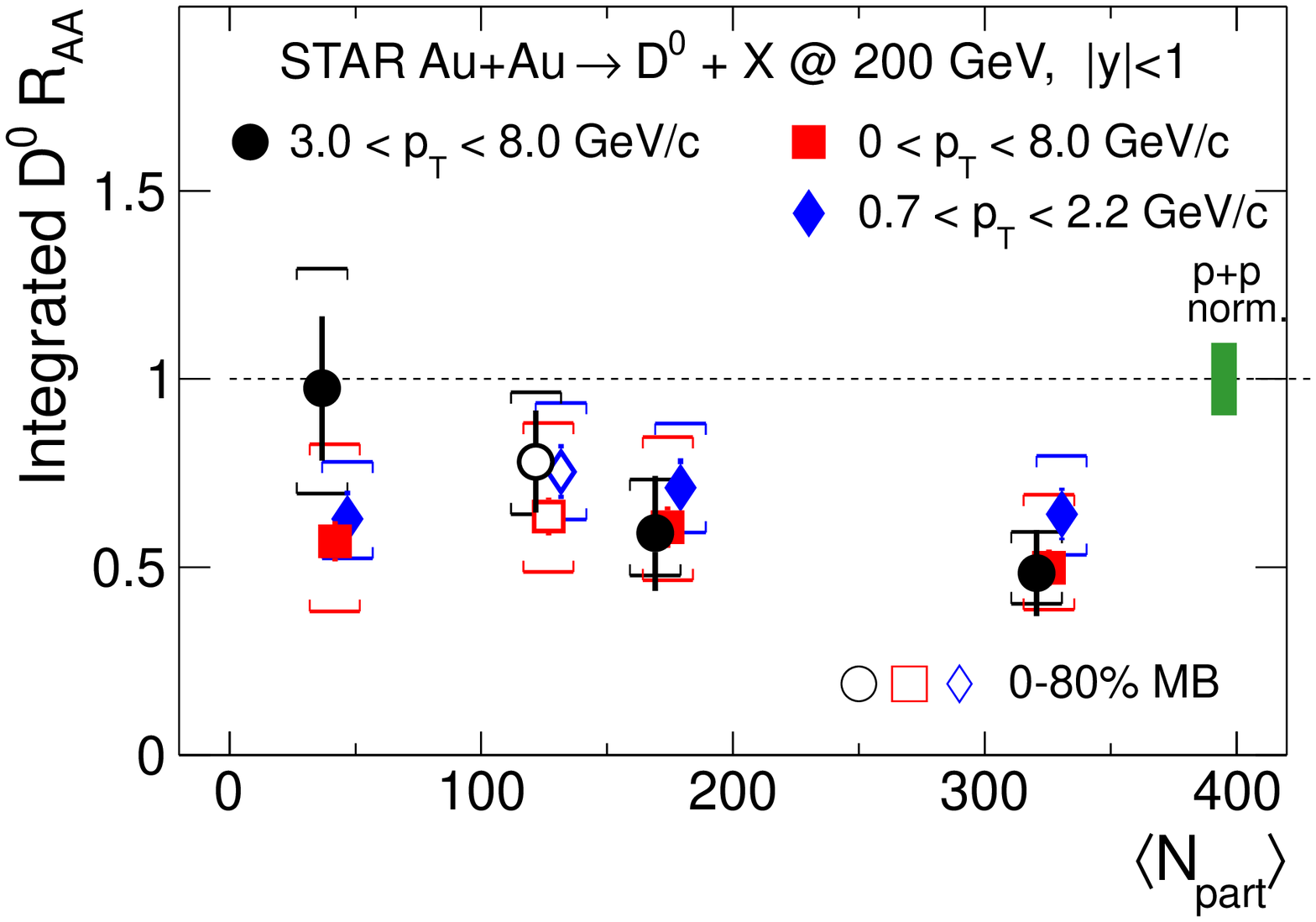}
\caption{(Color online) Integrated \dzero\ \RAA\  as a function of \npart\ in different \pt\ regions: 0$-$8 \gevc\ (squares), 0.7$-$2.2 \gevc\ (diamonds) and 3$-$8 \gevc\ (circles). 
Open symbols are for the 0$-$80\% minimum bias events. The vertical bar around unity denotes the overall normalization uncertainty from \pp\ reference.
} 
\label{fig:figure3}
\end{figure}

After correcting the two errors, the new \dzero\ \pt\ spectra are shown in Fig.~\ref{fig:figure1} as solid symbols for different centrality bins. The vertical bars on the points represent the statistical uncertainties and the brackets denote the systematic uncertainties. The measured \dzero\ production cross section per nucleon-nucleon-collision at mid-rapidity in the 0\%$-$10\% most-central collisions is updated as 41 $\pm$ 4(stat) $\pm$ 5(syst) $\mu$b. The \pp\ $D^0$ reference spectrum, shown as open circles, was obtained using the \dzero\ measurement at \pt\ $<$ 2.0 \gevc\ and $D^{*{\pm}}$ measurement at \pt\ $>$ 2.0 \gevc. 
We updated our $p+p$ reference spectrum in this erratum using the latest global analysis of charm fragmentation ratios from Ref.~\cite{Charmfrag} and also by taking into account the \pt\ dependence of the fragmentation ratio between \dzero\ and $D^{*{\pm}}$ from PYTHIA 6.4~\cite{PY6}, which increases the yield as \pt\ increases, reaching 40\%. The dashed curves are Levy function~\cite{Levy} fits to the \pp\ reference, scaled by the number of binary collisions, \nbin ~\cite{starsys}. In our previous \pp\ collision analysis, TOF matching was always required for the entire $p_T$ region in order to minimize the pile-up impact in high luminosity \pp\ collisions. Therefore, the \pp\ data were not affected by the error in the Au+Au analysis due to a less complicated algorithm. With these new updates, the \pp\ \dzero\ production cross section at mid-rapidity is measured as 80 $\pm$ 11(stat) $\pm$ 16(syst) $\mu$b.

Figure~\ref{fig:figure2} shows \dzero\ \RAA\ for the centrality bins of 40$-$80\% (a), 10$-$40\% (b) and 0$-$10\% (c). The vertical lines and brackets indicate the statistical and systematic uncertainties, respectively. The vertical bars around unity from left to right represent the uncertainties for \nbin\ and \pp\ cross-section, respectively. 
The \dzero\ \RAA\ as a function of \pt\ is calculated as the ratio between the \dzero\ yield in each \pt\ bin for each centrality of \AuAu\ collisions to the Levy function fit to the \pp\ data scaled by \nbin ~\cite{ppCharm}. 
The statistical and systematic uncertainties of the $p+p$ reference are displayed in this figure only within the systematic uncertainty of $R_{AA}$.  The uncertainty in the $p+p$ reference dominates this systematic uncertainty, and includes the 1 $\sigma$ uncertainty from the Levy fit and the difference between Levy and power-law function fits for extrapolation to low and high \pt, expressed as 1 standard deviation.
The conclusion of strong suppression observed in 0$-$10\% central collisions for \pt\ $>$ 2.5 \gevc\ still holds, while it is consistent with unity in peripheral collisions in this $p_T$ region. At \pt\ $<$ 1 \gevc, the $D^0$ yield is found to be suppressed in all centralities. The total charm quark pair yield is expected to follow \nbin\ scaling as charm quarks are believed to be predominately produced in initial hard scatterings. Charm quark hadronization from a coalescence mechanism may lead to an enhancement in the relative fractions of $D_s$ and $\Lambda_c$ hadrons in heavy-ion collisions~\cite{DsLcModel}, therefore resulting in a reduction in the observed \dzero\ yields in Au+Au collisions. In addition, cold nuclear matter effects, e.g. nuclear shadowing effect in gluon parton distributions, may also play an important role. 
In 0$-$10\% collisions, the suppression level is around 0.5 for \pt\ $>$ 3 \gevc , which is comparable to both the measurements of electrons from heavy flavor hadron decays~\cite{starcharmraa, phenixcharmraa} and the light hadrons~\cite{LFraa}. 

The integrated \RAA\ over \pt\ is calculated as a ratio of the integrated \dzero\ yield in Au+Au collisions to that of the \pp\ reference scaled by the \nbin\ in the given \pt\ region. Figure~\ref{fig:figure3} shows the integrated \dzero\ \RAA\ as a function of number of participants (\npart), which represents the collision centrality from the Glauber model~\cite{glauber}. The \RAA\ for 0 $<$ $p_T$ $<$ 8 \gevc\ is suppressed in all centralities and exhibits a weak dependence on \npart. 
The integrated \RAA\ of \dzero\ is more suppressed at high \pt\ in more central collisions.

In summary, the original conclusion in the published paper~\cite{D0PRL} about the suppression of the \dzero\ \RAA\ at \pt\ $>$ 3 \gevc\ is still valid. The bump structure in the intermediate \pt\ region is still there but no significant enhancement is observed. Since the \dzero\ cross section is suppressed integrated over all \pt, it is difficulty to draw a conclusion on the binary scaling of the total charm production cross section, which requires other charmed hadron measurements.

We thank the RHIC Operations Group and RCF at BNL, the NERSC Center at LBNL, and the Open Science Grid consortium for providing resources and support.  This work was supported in part by the Office of Nuclear Physics within the U.S. DOE Office of Science, the U.S. National Science Foundation, the Ministry of Education and Science of the Russian Federation, National 
Natural Science Foundation of China, Chinese Academy of Science, the Ministry of Science and Technology of China and the Chinese Ministry of Education, the National Research Foundation of Korea, Czech Science Foundation and Ministry of Education, Youth and Sports of the Czech Republic, Department of Atomic Energy and Department of Science and Technology of the Government of India, the National Science Centre of Poland, the Ministry  of Science, Education and Sports of the Republic of Croatia, RosAtom of Russia and German Bundesministerium fur Bildung, Wissenschaft, Forschung and Technologie (BMBF) and the Helmholtz Association.

\end{document}